\documentclass[prl, twocolumn, superscriptaddress, showpacs, preprintnumbers, amsmath, amssymb]{revtex4} 

\usepackage{graphicx}
\usepackage{epstopdf}
\usepackage{mathrsfs}
\usepackage{pifont}

\newcommand{\field}{h}
\newcommand{\pop}{n}
\newcommand{\Nedge}{\mathscr{N}}
\newcommand{\rate}{\nu}
\newcommand{\ud}{\mathrm{d}}

\newcommand{\vone}{\hbox{\large\ding{172}}}
\newcommand{\vtwo}{\hbox{\large\ding{173}}}
\newcommand{\vthree}{\hbox{\large\ding{174}}}

\begin{document}

\title{Vertex dynamics in finite two dimensional square spin ices}

\author{Zoe Budrikis}
\affiliation{School of Physics M013, University of Western Australia, 35 Stirling Hwy, Crawley WA 6009,
Australia}
\affiliation{Istituto dei Sistemi Complessi, Consiglio Nazionale delle Ricerche, Via Madonna del Piano
10, 50019 Sesto Fiorentino, Italy}
\author{Paolo Politi}
\affiliation{Istituto dei Sistemi Complessi, Consiglio Nazionale delle Ricerche,
Via Madonna del Piano 10, 50019 Sesto Fiorentino, Italy}
\affiliation{INFN Sezione di Firenze, via G. Sansone 1, 50019 Sesto Fiorentino, Italy}
\author{R. L. Stamps}
\affiliation{School of Physics M013, University of Western Australia, 35 Stirling Hwy, Crawley WA 6009,
Australia}

\begin{abstract}
Local magnetic ordering in artificial spin ices is discussed from
the point of view of how geometrical frustration controls dynamics
and the approach to steady state. We discuss the possibility of
using a particle picture based on vertex configurations to interpret
time evolution of magnetic configurations. Analysis of possible vertex processes
allows us to anticipate different behaviors for open and closed edges and the
existence of different field regimes. Numerical simulations confirm these
results and also demonstrate the importance of correlations
and long range interactions in understanding particle population
evolution. We also show that a mean field model of vertex dynamics gives important insights into finite size effects.
\end{abstract}


\maketitle

{\it Introduction.}---
Spin ices are geometrically frustrated magnetic systems, where interaction energies
are minimized by local arrangements of spins resembling the ice rule for the
ground state of solid water, i.e., two-in, two-out spin configurations \cite{Harris:1997, Bramwell:2001}.
Spin ices display several interesting features including zero point entropy \cite{Ramirez:1999}, spin freezing, and hysteresis~\cite{freezing}. The role of long range interactions in three dimensional spin ices is also interesting, as the long range dipolar interactions between spins lead to a state that can
be described using the short range ice rules \cite{Isakov:2005}.

There is now a growing interest in two dimensional artificial spin
ices, which consist of finite arrays of elongated magnetic dots whose magnetizations can be well approximated by Ising spins \cite{Wang:2006, Qi:2008}. Artificial spin ices have several differences with their three dimensional counterparts. For example, their geometry can be controlled experimentally \cite{Li:2010} and their magnetization configurations can be imaged directly using scanning probe techniques.
Furthermore, artificial spin ices allow the study of frustration at room temperature because of the thermal stability of the relatively large dots.
In fact, the anisotropy barrier preventing the macrospin
associated with a dot from flipping is much larger than room temperature, so dynamics can only be induced by applying a magnetic field $\vec H= h\hat\theta$. We can expect that a small $h$
does not affect the configuration, while a large $h$ dominates dipolar interactions
and spins simply follow the field \cite{Ke:2008}. Interesting collective effects arise when $|h-h_c| \sim g$, where $h_c$ is the minimum field required to flip an isolated spin and $g$ is of the order of the dipolar interactions.

Until now, studies of artificial spin ices have focused on the best procedures to
attain low energy states \cite{Ke:2008} or the properties of `final' states
of demagnetization protocols \cite{Nisoli:2007, Nisoli:2010}.
Very little is known about the time-dependent dynamical evolution
of spin ices in response to a magnetic field, either experimentally or theoretically.
In this Letter we show that the time evolution of square artificial spin ices can be described in terms of vertex configurations, and that these interact in a way that in principle can be predicted and understood.

We do this by first introducing the essentials of a vertex population model,
which predicts some general features of the dynamics. We show that a mean field approximation
allows quantitative evaluation of important array size effects.
However, correlations, the long range part of dipolar interactions,
and edge effects can be nonnegligible and are  studied carefully using
numerical simulations.
In all of our models, a rotating field of constant strength is applied to finite arrays and we discuss how this leads to effective demagnetization if $h$ is correctly tuned.

\begin{figure}
  \includegraphics[width=\columnwidth]{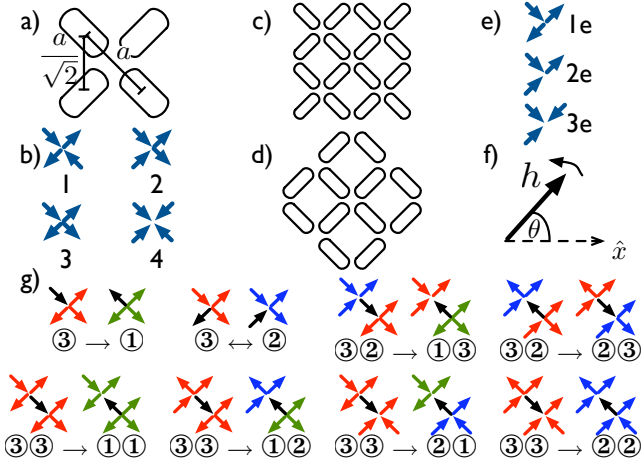}
  \caption{\label{array_geometries}The geometry of square artificial spin ice. (a) The nearest and next nearest neighbor interactions at each 4-island vertex of the array; (b) examples of each type of 4-island vertex; (c) `open' edge geometry which can be described fully by the 4-island vertex configuration; (d) `closed' edge geometry with vertices containing three islands; (e) examples of each type of 3-island vertex; (f) the magnitude and direction of the applied magnetic field; (g) the single spin flip one- and two-vertex processes that are energetically and topologically allowed. The spin that flips is black, while spins in type 1, 2, and 3 vertices are green, blue and red, respectively.}
\end{figure}

{\it Vertex population model.}---
In this work, we consider only square artificial spin ices. Unlike the spins on the tetrahedra of three
dimensional ices, the four spins around a vertex of a square ice are not
equivalent (Fig. \ref{array_geometries}a). As in Ref. \cite{Wang:2006}, we classify the vertex configurations into four types of increasing energy
$E_i$ ($i=1,\dots,4$) (Fig. \ref{array_geometries}b).
Type 1 and 2 vertices obey the two-in, two-out ice rule but
$E_1<E_2$, because type 1 vertices
minimize the energy of stronger interactions.

The arrays can have open or closed edges (Fig. \ref{array_geometries}c,d).
Unlike in the open edge case,
the configuration of an array with closed edges cannot be fully described by the configuration of its 4-island vertices and three types of 3-island edge vertices must be introduced (Fig. \ref{array_geometries}e).

We now discuss the essentials of a vertex population model for open edge arrays. After sample saturation in the $\hat x$ direction, a field of constant modulus $h$ is applied and rotated
anticlockwise (Fig. \ref{array_geometries}f). The system evolves via single spin flips rather than so-called `loop moves' \cite{Barkema:1998} so that ice-rule disobeying vertices can arise, a necessary condition to describe dynamics.
Let us suppose that the condition for flipping a spin $\vec s_i$ is
$-\vec H^{\mathrm{tot}}_i\cdot\vec s_i > h_c$. $\vec H^{\mathrm{tot}}_i$ is the total field acting on $\vec s_i$, and is the sum of the external field $\vec H$ and the
dipolar field $\vec H^{\mathrm{dip}}_i$ due to all other spins. 
We work in reduced units where the energy of a nearest-neighbor pair of spins is $\pm 3/2$, and set $h_c=10$, which is larger than the largest possible dipolar field $\tilde{g}$ acting on any island \cite{note_hc}. As long as $h_c > \tilde{g}$, results with different $h_c$ are simply shifted relative to one another.

We introduce the notation that \textcircled{$i$} refers to a type $i$ vertex, and $\pop_i$ the population fraction of type $i$ vertices. In the bulk of the array, topological and energetic constraints allow the following 2-vertex processes:
\begin{gather*}
\vthree\vtwo \to \vone\vthree, \vtwo\vthree \\
\vthree\vthree \to \vone\vone, \vone\vtwo, \vtwo\vone, \vtwo\vtwo
\end{gather*}
as shown in Fig. \ref{array_geometries}g.
Because of the energetic constraints, type 4 vertices cannot be created from an initial polarised state,
type 1 vertices cannot be destroyed, and the process $\vtwo\vtwo\to \vthree\vthree$ cannot occur. The impossibility of destroying type 1 vertices and the impossibility
of the process $\vtwo\vtwo\to \vthree\vthree$ give rise to an important phenomenon, trapping, in which a type 3 vertex or a region of type 2 vertices is frozen if surrounded by type 1 vertices. The existence of trapping contributes to a slowing of dynamics at long times. Type 3 vertices can only be nucleated and expelled at the array edges, via the processes $\vthree \to \vone$ and  $\vthree \leftrightarrow \vtwo$, also shown in Fig. \ref{array_geometries}g.

Starting from an initial saturated configuration, the dynamics begin with the nucleation of type 3 vertices at the array edges perpendicular to the original polarization, via the process $\vtwo\to\vthree$.
Then type 3 vertices can propagate via two processes, $\vthree\vtwo \to \vone\vthree$ and $\vthree\vtwo \to \vtwo\vthree$, the former requiring a smaller applied field than the latter.

In the absence of long-range interactions, there would be four field regimes: a very low field regime where the field is too small to effect a response from the spins, a very high field regime in which the magnetization tracks the applied field, and two nontrivial intermediate regimes. The difference between the two intermediate regimes is that in the first (the `low field' regime), type 3 vertices can only propagate via the process $\vthree\vtwo \to \vone\vthree$ whereas in the second (the `high field' regime), both type 3 propagation processes can occur. Long range interactions modify this picture somewhat because the dipolar fields acting on spins in different locations are not equal, causing a spread in the applied field strength required for vertex processes. This leads to a more gradual crossover between the low and high field regimes.

A full description of vertex dynamics should take into account correlations and long range interactions, but if vertices are sufficiently well-mixed then these should become negligible on average and a mean field description based on short range interactions should hold. We propose a set of ordinary differential equations to describe an open edge array under these assumptions. Comparison with numerical simulations that take into account the full long range interactions (see below) shows that the mean field approach describes the crossover and high field regimes relatively well.

The equations for the $n_i(\theta)$ are:
\begin{subequations}
\label{pde_open}
\begin{align}
\label{open_type1}
\begin{split}
\dot{\pop}_1 &=	4\rate_{13}  (1-\Nedge/2) \pop_{2F} \pop_{3F}\\ 
	& + 2(2\nu_{11}+\nu_{12}) (1-\Nedge/2) \pop_{3F}^2 
	+ \nu_{31}^e \Nedge \pop_{3F}, 
\end{split}\\
\begin{split}
\dot{\pop}_{2F} &=	-4\rate_{13} (1-\Nedge/2) \pop_{2F} \pop_{3F}\\  
	& + 2(\nu_{12}+2\nu_{22}) (1-\Nedge/2) \pop_{3F}^2 
	-\nu_{23}^e \Nedge \pop_{2F}\\ 
	& + \nu_{32}^e \Nedge \pop_{3F} 
	 - \rate_T (1-\Nedge) \pop_{2F} \pop_1^4, 
\end{split}\\
\begin{split}
\dot{\pop}_{3F} &=	-4(\nu_{11}+\nu_{12}+\nu_{22}) (1-\Nedge/2) \pop_{3F}^2 
	+ \nu_{23}^e \Nedge \pop_{2F} \\ 
	&- (\nu_{31}^e+\nu_{32}^e) \Nedge \pop_{3F} 
	- \rate_T (1-\Nedge) \pop_{3F} \pop_1^3,
\end{split}\\
\dot{\pop}_{iT} &= \rate_T (1-\Nedge) \pop_{iF} \pop_1^4 \qquad (i=2,3), 
\end{align}
\end{subequations}
where $\dot n_i=\ud n_i/\ud \theta$, $\Nedge$ is the fraction of vertices at the array edge, and subscripts `T' and `F' denote vertices that are trapped and vertices that are free to be involved in dynamics, respectively. 
The reaction rates $\nu$ depend on the total field acting on the flipping spin.
The dipolar part is calculated in the approximation that only spins
belonging to the spin's two vertices contribute to it.

\begin{figure}
  \includegraphics[width=\columnwidth]{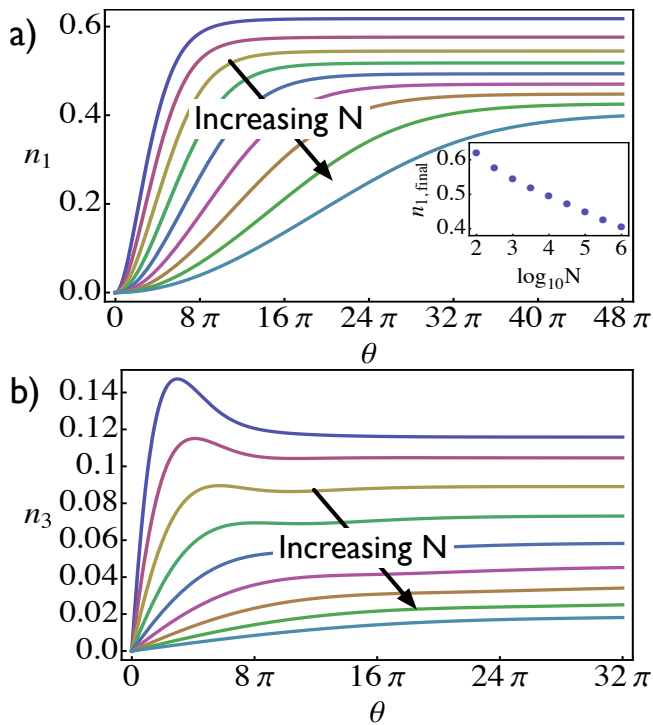}
  \caption{\label{diagonal_model_populations}(a) Type 1 and (b) type 3 population fractions for dynamics given by Eqs. \eqref{pde_open}, for a field $h=10$ and for arrays of increasing size. Inset: Final type 1 population fraction as a function of array size $N$.}
\end{figure}

We integrate the system \eqref{pde_open} with initial conditions $\pop_{2F}(0) = 1, \pop_i(0)=0$ for all other $i$, for $h=10$.
In Fig. \ref{diagonal_model_populations}, we see that size effects are clearly significant for population levels: as the number of vertices $N$ increases the final type 1 and 3 populations decrease logarithmically slowly (see Inset), the number of rotations required to reach a steady state increases, and the transient peak in $\pop_3$ disappears. These effects can all be attributed to the decrease in $\Nedge$ with $N$, because for large arrays with relatively few edge vertices, the rate of type 3 vertex creation will be smaller, resulting in reduced rates for processes that create type 1 vertices.

Considerations for closed edge arrays can be made in direct analogy to those presented above for open edge arrays. The resulting equations are more complicated due to the additional populations introduced by the edge geometry, as noted earlier, but the resulting behaviors for $\pop_1$ and $\pop_3$  turn out to be very similar to those shown in Fig. \ref{diagonal_model_populations}.  This similarity signals a failure of our mean field approximation.

{\it Correlations and long range dipolar effects.}---
The primary shortcoming of our mean field population dynamics picture is the neglect 
of spin correlations, but the neglect of long range dipolar interactions is also important.
We demonstrate this using numerical simulations which do not suffer from these two shortcomings. At each step of a simulation, all spins are accessed in a random order, and flipped according to the $h_c$ rule described above. The field is held constant until no further spin flips can occur and then rotated by $d\theta=0.01$. The initial configuration is saturated in the $\hat{x}$ direction. Example animated time evolutions for open and closed arrays can be found in the Supplementary Materials \cite{EPAPS_Document}.

\begin{figure}
  \includegraphics[width=\columnwidth]{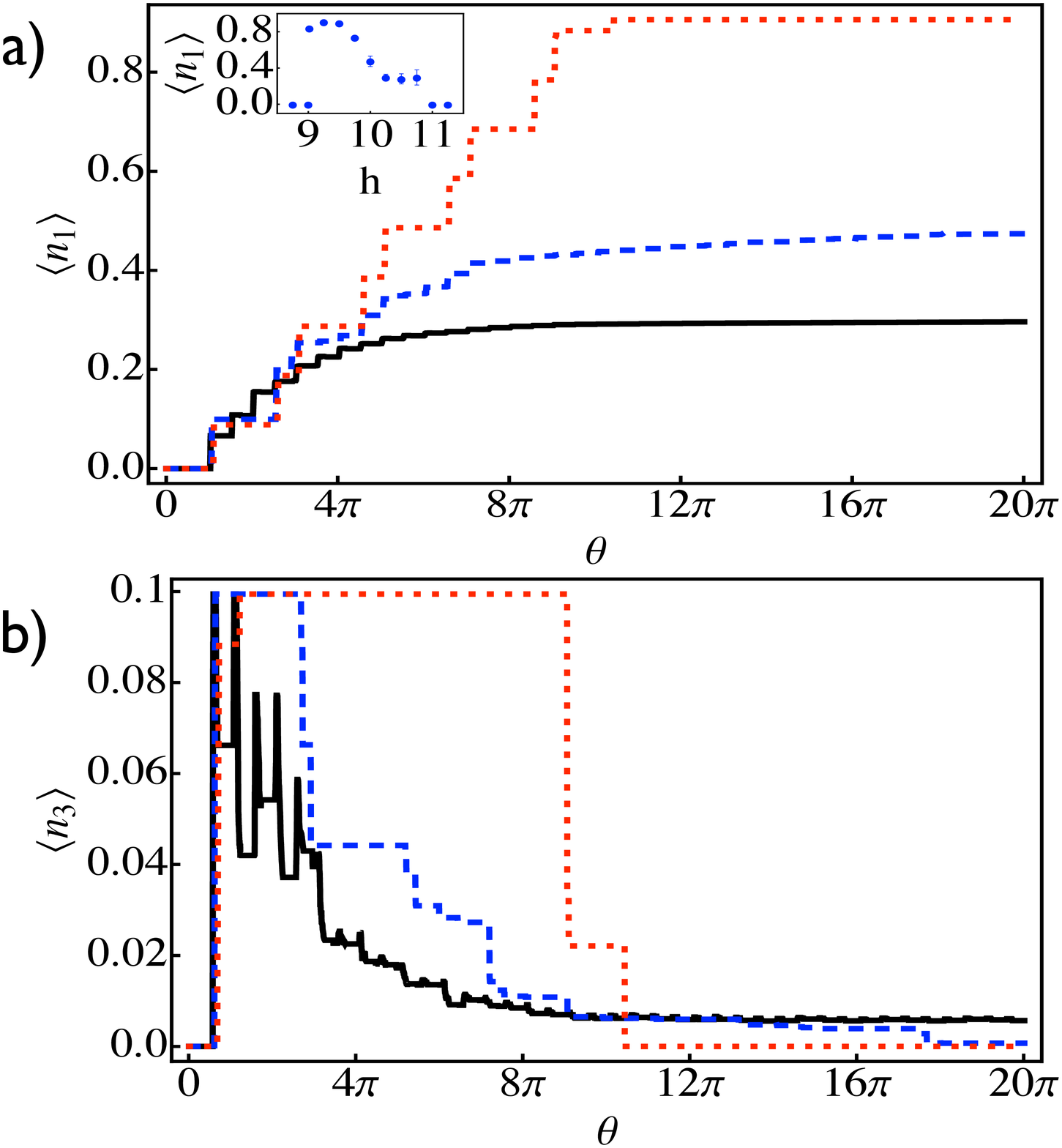}
  \caption{\label{diagonal_simulation_populations}(a) Type 1 and (b) type 3 mean vertex populations from numerical simulations, for $\field=10.75$
(solid black lines), $10$ (blue dashed lines), and $9.25$ (red dotted lines), for an open edge array of 400 islands. Averages are over 100 runs of the simulation. The inset of (a) shows the
dependence of the final type 1 population on $\field$, with the two field regimes of interest: low fields ($9 \le \field \le 9.5$) and high fields ($10 \le \field \le 11$).}
\end{figure}

In Fig. \ref{diagonal_simulation_populations} we plot
$\langle\pop_1(\theta)\rangle$ and $\langle\pop_3(\theta)\rangle$ for an open edge array. The inset shows the dependence of the final type 1 population fraction on the field strength $h$.
We find four field regimes, as expected by our general considerations of vertex
dynamics. The lowest and highest regimes are trivial: for $h\le 9$, no spin
flips can occur, while for $h\ge 11$ spins simply follow the field. In the interval $9\le h\le 11$, the low field regime
(red dotted lines) and high field regime (solid black lines) are shown,
together with a crossover between the two (dashed blue lines).
It is worth stressing
we are able to attain $\langle\pop_1\rangle$ values of up to 90\% if the field is slightly larger than the lowest field allowing dynamics.

\begin{figure}
  \includegraphics[width=\columnwidth]{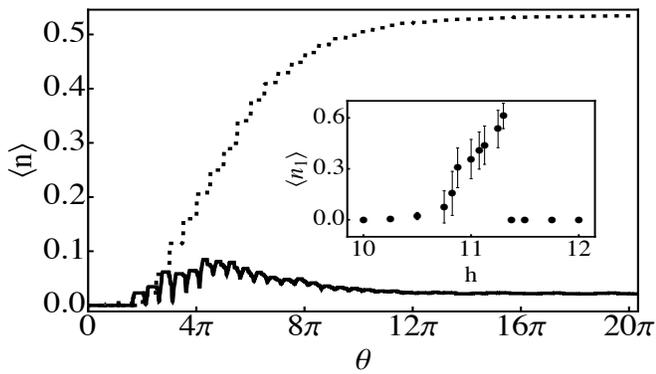}
  \caption{\label{closed_simulation_populations}Type 1 (dotted line) and type 3 (solid line) vertex population fractions from simulations for a closed edge array of 220 islands with an applied field
$\field=11.25$. Averages are over 100 runs of the simulation. Inset: Dependence of the mean final type 1 population fraction on field strength $h$.}
\end{figure}

We performed similar simulations for closed edge arrays. As an example of the results, Fig. \ref{closed_simulation_populations} shows the evolution of mean $\pop_1$ and $\pop_3$ values for an applied field $h=11.25$, with an inset showing the dependence of the final $\langle\pop_1\rangle$ on $h$. These results are quite different to those obtained for open edge array for several reasons.

The same 2-vertex processes occur in the bulk of open and closed arrays and likewise type 3 nucleation and expulsion occurs at the edges of closed arrays.
However, in closed edge arrays, type 3 nucleation and expulsion can only occur when the neighboring 3-island edge vertex is of type $2e$, because the fields required when it is type $1e$ are prohibitively large. Thus, creation of type $1e$ vertices reduces
the number of nucleation and expulsion sites and slows the dynamics.

Moreover the fields required to nucleate type 3 vertices in the closed edge array
are larger than the field required for the
$\vthree\vtwo\to\vtwo\vthree$ process, so this process is always able to occur.
As a result, there is only one interesting field regime, for $10.5<h < 11.5$.
The field required for initial type 3 nucleation is sufficiently large that a relatively high proportion of nucleated type 3 vertices are driven across the array in the process $\vthree\vtwo\to\vtwo\vthree$ and expelled at the opposite edge. This leads to oscillations in $\langle\pop_3\rangle$.

{\it Conclusions.}---
We have demonstrated in this paper that vertex dynamics can be formalized via population dynamics equations. We have done this within the limitations of a short range interaction mean field approximation, and provided
insights into array size dependence. Most significantly, we find that the final number of type 1 vertices decreases very slowly with increasing array size.

It may be possible to improve the mean field treatment
 by introducing spatially dependent densities for the populations and estimating effects of spin correlations on transition rates, but that is beyond the scope of the present work. However, we note that a mean field approximation can provide an accurate description in some circumstances. The reason is the following. While evolution begins with nucleation of type 3 vertices at the array edge, numerically we find that different edges produce very different results. Instead if the initial configuration is random
rather than saturated, many defects are included and edge effects
become negligible. Thus a random initial configuration is described well in a mean field approach.
 
Based on results from our models, we can also make a general observation of importance for the question of demagnetization protocol. We find that a simple constant $h$ protocol may be as good as
more elaborate protocols, if $h$ is correctly tuned. We have also studied the system using some
protocols that change both $\theta$ and $h$. The best demagnetization appears when $h$ is varied such that
the system spends a long time with $h$ near the values that give the maximum mean $\pop_1$ in the inset of Fig. \ref{diagonal_simulation_populations}a.

Lastly, we also note that real islands may be modeled more accurately with Stoner-Wohlfarth switching. We have examined this and find that we obtain qualitatively similar results for critical field and Stoner-Wohlfarth switching, but critical field switching is more amenable to a simple theoretical description.

\begin{acknowledgments}
Z.B. and R.L.S. thank the ARC and the WUN for support. Z.B. also acknowledges ARNAM and the Hackett Foundation.
\end{acknowledgments}


\begin{thebibliography}{1}%
\makeatletter
\providecommand \@ifxundefined [1]{%
 \ifx #1\undefined \expandafter \@firstoftwo
 \else \expandafter \@secondoftwo
\fi
}%
\providecommand \@ifnum [1]{%
 \ifnum #1\expandafter \@firstoftwo
 \else \expandafter \@secondoftwo
\fi
}%
\providecommand \enquote [1]{``#1''}%
\providecommand \bibnamefont  [1]{#1}%
\providecommand \bibfnamefont [1]{#1}%
\providecommand \citenamefont [1]{#1}%
\providecommand\href[0]{\@sanitize\@href}%
\providecommand\@href[1]{\endgroup\@@startlink{#1}\endgroup\@@href}%
\providecommand\@@href[1]{#1\@@endlink}%
\providecommand \@sanitize [0]{\begingroup\catcode`\&12\catcode`\#12\relax}%
\@ifxundefined \pdfoutput {\@firstoftwo}{%
 \@ifnum{\z@=\pdfoutput}{\@firstoftwo}{\@secondoftwo}%
}{%
 \providecommand\@@startlink[1]{\leavevmode}%
 \providecommand\@@endlink[0]{}%
}{%
 \providecommand\@@startlink[1]{%
  \leavevmode
  \pdfstartlink
   attr{/Border[0 0 1 ]/H/I/C[0 1 1]}%
   user{/Subtype/Link/A<</Type/Action/S/URI/URI(#1)>>}%
  \relax
 }%
 \providecommand\@@endlink[0]{\pdfendlink}%
}%
\providecommand \url  [0]{\begingroup\@sanitize \@url }%
\providecommand \@url [1]{\endgroup\@href {#1}{\urlprefix}}%
\providecommand \urlprefix [0]{URL }%
\providecommand \Eprint[0]{\href }%
\@ifxundefined \urlstyle {%
  \providecommand \doi [1]{doi:\discretionary{}{}{}#1}%
}{%
  \providecommand \doi [0]{doi:\discretionary{}{}{}\begingroup
  \urlstyle{rm}\Url }%
}%
\providecommand \doibase [0]{http://dx.doi.org/}%
\providecommand \Doi[1]{\href{\doibase#1}}%
\providecommand \bibAnnote [3]{%
  \BibitemShut{#1}%
  \begin{quotation}\noindent
    \textsc{Key:}\ #2\\\textsc{Annotation:}\ #3%
  \end{quotation}%
}%
\providecommand \bibAnnoteFile [2]{%
  \IfFileExists{#2}{\bibAnnote {#1} {#2} {\input{#2}}}{}%
}%
\providecommand \typeout [0]{\immediate \write \m@ne }%
\providecommand \selectlanguage [0]{\@gobble}%
\providecommand \bibinfo [0]{\@secondoftwo}%
\providecommand \bibfield [0]{\@secondoftwo}%
\providecommand \translation [1]{[#1]}%
\providecommand \BibitemOpen[0]{}%
\providecommand \bibitemStop [0]{}%
\providecommand \bibitemNoStop [0]{.\EOS\space}%
\providecommand \EOS [0]{\spacefactor3000\relax}%
\providecommand \BibitemShut [1]{\csname bibitem#1\endcsname}%
%</preamble>
\bibitem{Harris:1997}%
  \BibitemOpen
  \bibfield{author}{%
  \bibinfo {author} {\bibfnamefont{M.~J.}\ \bibnamefont{Harris}}, \bibinfo
  {author} {\bibfnamefont{S.~T.}\ \bibnamefont{Bramwell}}, \bibinfo {author}
  {\bibfnamefont{D.~F.}\ \bibnamefont{McMorrow}}, \bibinfo {author}
  {\bibfnamefont{T.}~\bibnamefont{Zeiske}},\ and\ \bibinfo {author}
  {\bibfnamefont{K.~W.}\ \bibnamefont{Godfrey}},\ }%
  \bibfield{journal}{%
	{\bibinfo {journal} {Phys. Rev. Lett.}}\ }%
  \textbf{\bibinfo {volume} {79}},\ \bibinfo {pages} {2554} (\bibinfo {year} {1997}).%
\bibitem{Bramwell:2001}%
  \BibitemOpen
  \bibfield{author}{%
  \bibinfo {author} {\bibfnamefont{S.~T.}\ \bibnamefont{Bramwell}}\ and\
  \bibinfo {author} {\bibfnamefont{M.~J.~P.}\ \bibnamefont{Gingras}},\ }%
  \bibfield{journal}{%
  \bibinfo {journal} {Science}\ }%
  \textbf{\bibinfo {volume} {294}},\ \bibinfo {pages} {1495} (\bibinfo {year} {2001}).\
\bibitem{Ramirez:1999}%
  \BibitemOpen
  \bibfield{author}{%
  \bibinfo {author} {\bibfnamefont{A.~P.}\ \bibnamefont{Ramirez}}, \bibinfo
  {author} {\bibfnamefont{A.}~\bibnamefont{Hayashi}}, \bibinfo {author}
  {\bibfnamefont{R.~J.}\ \bibnamefont{Cava}}, \bibinfo {author}
  {\bibfnamefont{R.}~\bibnamefont{Siddharthan}},\ and\ \bibinfo {author}
  {\bibfnamefont{B.~S.}\ \bibnamefont{Shastry}},\ }%
  \bibfield{journal}{%
  \bibinfo {journal} {Nature}\ }%
  \textbf{\bibinfo {volume} {399}},\ \bibinfo {pages} {333} (\bibinfo {year}
  {1999}).
\bibitem{freezing}
J. Snyder, B. G. Ueland, J. S. Slusky, H. Karunadasa, R. J. Cava, and P. Schiffer,
Phys. Rev. B {\bf 69}, 064414 (2004).
\bibitem{Isakov:2005}%
  \BibitemOpen
  \bibfield{author}{%
  \bibinfo {author} {\bibfnamefont{S.~V.}\ \bibnamefont{Isakov}}, \bibinfo
  {author} {\bibfnamefont{R.}~\bibnamefont{Moessner}},\ and\ \bibinfo {author}
  {\bibfnamefont{S.~L.}\ \bibnamefont{Sondhi}},\ }%
  \bibfield{journal}{%
  {\bibinfo {journal} {Phys. Rev. Lett.}}\ }%
  \textbf{\bibinfo {volume} {95}}, \bibinfo{pages} {217201} (\bibinfo {year}
  {2005}).
\bibitem{Wang:2006}%
R. F. Wang {\it et al.}, Nature {\bf 439}, 303 (2006).
\bibitem{Qi:2008}%
  \BibitemOpen
  \bibfield{author}{%
  \bibinfo {author} {\bibfnamefont{Y.}~\bibnamefont{Qi}}, \bibinfo {author}
  {\bibfnamefont{T.}~\bibnamefont{Brintlinger}},\ and\ \bibinfo {author}
  {\bibfnamefont{J.}~\bibnamefont{Cumings}},\ }%
  \bibfield{journal}{%
  {\bibinfo {journal} {Phys. Rev. B}}\ }%
  \textbf{\bibinfo {volume} {77}}, \bibinfo{pages}{094418} (\bibinfo {year}
  {2008}).
\bibitem{Li:2010}%
  \BibitemOpen
  \bibfield{author}{%
  \bibinfo {author} {\bibfnamefont{J.}~\bibnamefont{Li}}, \bibinfo {author}
  {\bibfnamefont{X.}~\bibnamefont{Ke}},\ \bibinfo {author}
  {\bibfnamefont{S.}~\bibnamefont{Zhang}},\ \bibinfo {author}
  {\bibfnamefont{D.}~\bibnamefont{Garand}},\ \bibinfo {author}
  {\bibfnamefont{C.}~\bibnamefont{Nisoli}},\ \bibinfo {author}
  {\bibfnamefont{P.}~\bibnamefont{Lammert}},\ \bibinfo {author}
  {\bibfnamefont{V. H.}~\bibnamefont{Crespi}},\ and \bibinfo {author}
  {\bibfnamefont{P.}~\bibnamefont{Schiffer}},\ }%
  \bibfield{journal}{%
  {\bibinfo {journal} {Phys. Rev. B}}\ }%
  \textbf{\bibinfo {volume} {81}}, \bibinfo{pages}{092406} (\bibinfo {year}
  {2010}).
\bibitem{Ke:2008}%
  \BibitemOpen
  \bibfield{author}{%
  \bibinfo {author} {\bibfnamefont{X.}~\bibnamefont{Ke}}, \bibinfo {author}
  {\bibfnamefont{J.}~\bibnamefont{Li}}, \bibinfo {author}
  {\bibfnamefont{C.}~\bibnamefont{Nisoli}}, \bibinfo {author}
  {\bibfnamefont{P.~E.}\ \bibnamefont{Lammert}}, \bibinfo {author}
  {\bibfnamefont{W.}~\bibnamefont{McConville}}, \bibinfo {author}
  {\bibfnamefont{R.~F.}\ \bibnamefont{Wang}}, \bibinfo {author}
  {\bibfnamefont{V.~H.}\ \bibnamefont{Crespi}},\ and\ \bibinfo {author}
  {\bibfnamefont{P.}~\bibnamefont{Schiffer}},\ }%
  \bibfield{journal}{%
  {\bibinfo {journal} {Phys. Rev. Lett.}}\ }%
  \textbf{\bibinfo {volume} {101}}, \bibinfo{pages}{037205} (\bibinfo {year}
  {2008}).
\bibitem{Nisoli:2007}%
  \BibitemOpen
  \bibfield{author}{%
  \bibinfo {author} {\bibfnamefont{C.}~\bibnamefont{Nisoli}}, \bibinfo {author}
  {\bibfnamefont{R.}~\bibnamefont{Wang}}, \bibinfo {author}
  {\bibfnamefont{J.}~\bibnamefont{Li}}, \bibinfo {author}
  {\bibfnamefont{W.~F.}\ \bibnamefont{McConville}}, \bibinfo {author}
  {\bibfnamefont{P.~E.}\ \bibnamefont{Lammert}}, \bibinfo {author}
  {\bibfnamefont{P.}~\bibnamefont{Schiffer}},\ and\ \bibinfo {author}
  {\bibfnamefont{V.~H.}\ \bibnamefont{Crespi}},\ }%
  \bibfield{journal}{%
  {\bibinfo {journal} {Phys. Rev. Lett.}}\ }%
  \textbf{\bibinfo {volume} {98}}, \bibinfo{pages}{217203} (\bibinfo {year}
  {2007}).
\bibitem{Nisoli:2010}
C. Nisoli, J. Li, X. Ke, D. Garand, P. Schiffer, and V. H. Crespi, arXiv:1005.3463 (unpublished).
\bibitem{Barkema:1998}%
  \BibitemOpen
  \bibfield{author}{%
  \bibinfo {author} {\bibfnamefont{G.~T.}~\bibnamefont{Barkema}} and \bibinfo {author} {\bibfnamefont{M.~E.~J.}~\bibnamefont{Newman}},\ }%
  \bibfield{journal}{%
  {\bibinfo {journal} {Phys. Rev. E}}\ }%
  \textbf{\bibinfo {volume} {57}}, \bibinfo{pages}{1155} (\bibinfo {year}
  {1998}).
\bibitem{note_hc}
In the array studied by Ke {\it et al.} \protect\cite{Ke:2008}, $h_c \approx 770$ Oe and $h^{\mathrm{dip}}\approx 10$ Oe.
\bibitem{EPAPS_Document}
See EPAPS Document X. 
For more information on EPAPS, see http://www.aip.org/pubservs/epaps.html.
\end{thebibliography}
\end{document}